\documentclass{amsart}
\usepackage{graphicx} 
\usepackage{lipsum}
\usepackage{url}
\usepackage{xcolor}

\title[De-novo Identification of Small Molecules from GC-EI-MS]{De-novo Identification of Small Molecules from Their GC-EI-MS Spectra}
\author[Adam Hájek et al.]{Adam Hájek \and Michal Starý \and Filip Jozefov \and Helge Hecht \and Elliott Price \and Aleš Křenek}
\date{March 2023}

\begin{document}

\begin{abstract}
Identification of experimentally acquired mass spectra of unknown compounds
presents a~particular challenge because reliable spectral databases do not cover
the potential chemical space with sufficient density.
Therefore machine learning based \emph{de-novo} methods, which derive molecular structure directly 
from its mass spectrum gained attention recently.
We present a~novel method in this family, addressing a~specific usecase of GC-EI-MS spectra,
which is particularly hard due to lack of additional information from the first stage of MS/MS
experiments, on which the previously published methods rely.
We analyze strengths and drawbacks or our approach and discuss future directions.
\end{abstract}

\maketitle

\section{Introduction}

Mass spectrometry is an analytical method constantly gaining importance in previous decades.
Besides well-established targeted or suspect screening modes (proving or reputing presence of a~compound),
the \emph{untargeted} mass spectrometry, i.e.~analyses of a~sample of unknown composition, 
is a~specific challenge.
Even if we restrict attention to ``small molecules'' (up to approx.~500 Da), the universe of possible compounds is huge,
despite that the well-known number of $10^{60}$~\cite{chemspace} can be overestimated towards combinatorially possible, not
really existing compounds.
In either case, the number largely supersedes the number of potentially synthesable compounds registered in ZINC~\cite{zinc}
database -- approx.~1,000,000,000.
Further, following the traditional approach of extracting a~series of per-compound mass spectra from experimental data, 
and searching for matches in databases of known spectra is even more restrictive.
The most comprehensive mass spectra library, NIST20, contains 350,704 spectra only.
Consequently, vast majority of compounds that can occur in experimental samples, are not reliably identifiable 
with spectral database search.

For this reason, together with breakthrough developments in machine learning techniques,
several \emph{de-novo} methods emerged recently, which are based on training a~model (artificial neural network)
with spectra on input and structural representation of the molecule (SMILES) on output.
The principal problem of relatively small size of spectral databases bounces back here -- the datasets 
are too small to be used to train the large models, which can grow up to 1,000,000,000 trainable parameters
trainable parameters easily.
The common pattern of these methods overcomes this lack of information by injecting much larger databases of compounds (ZINC)
and using these to generate spectra estimation.
Despite of certain inaccuracy, these inputs are sufficient to train the large model to perform quite well finally.
We describe several such methods in more detail in Sect.~\ref{background}.


In this work, we propose a~novel method to this family, which targets our GC-EI-MS usecase specifically.
According to our knowledge, no de-novo method directly applicable in this domain was developed before.

\section{Background}
\label{background}

Spec2Mol~\cite{spec2mol} is an encoder-decoder architecture, the encoder generates spectra embeddings 
and encoder reconstructs SMILES directly from them. 
The training is done in two stages separately, first the decoder (based on GRU NN architecture)
as a~part of SMILES-to-SMILES autoencoder using PubChem based training set of SMILES.
In the second stage, CNN-based spectra encoder is trained to end up with the same embeddings,
using experimental spectra from the NIST tandem spectral library.
It targets tandem (MS/MS) spectra, leveraging availability of precursor ion information, as well
multiple energy levels and [M+H]+ or [M-H]- ionization eventually.
For these dependencies the model is not directly applicable for our usecase.

SVAE~\cite{svae} uses variational autoencoder (VAE)-based architecture, which enhances the standard
autoencoder architecture with additional loss term derived from Kullback-Leibner 
divergence between latent space projection of a~training batch and normal distribution.
Similarly to Spec2Mol, two autoencoders (for spectra and SMILES) are trained simultaneously
to share the latent space, hence being able to generate mix encoder and decoder of both.
However, the implementation seems be rather experimental, and results on rather limited
datasets is reported only. 
Not much detail is revealed on the spectral part for us to be able to assess an intrinsic
dependency on tandem MS/MS again but all the results reported on these data.

MassGenie~\cite{massgenie} is the most inspiring work for us. 
First, a~forward model FragGenie, based on MetFrag~\cite{metfrag} is used to generate 
spectra estimation on 7.9 million SMILES drawn from ZINC. 
The forward model is based on semi-physical modeling, computing expected fragmentation patterns
based on more or less empirical rules.
In the second stage a transformer model is trained on the generated spectra.
Due to the use of FragGenie tightly tailored for ionization used in MS/MS the model is not usable
for our purposes as a~whole.

MSNovelist~\cite{msnovelist} splits the problem of de-novo identification in a~different way,
to a~spectrum-to-fingerprint and fingerprint-to-SMILES stages.
The former relies on SIRIUS~\cite{sirius} and CSI:FingerID~\cite{csifingerid},
the latter is a~LSTM network.
The common idea of larger structure-only training dataset comes here again; the fingerprint-to-SMILES
model is trained on such data.
Because both parts of the model use information on precursor ion heavily, the approach is not 
applicable for the GC-EI-MS usecase again.


NEIMS~\cite{neims} is a~relatively simple neural network predictor of EI mass spectra.
Its multilayer perceptron is trained to transform a~circular molecular fingerprint of lenght 4096 (computed in
a~standard way by RDKit) to spectrum represented as a~vector of intensities at individual $m/z$ values.
The resulting model trained on NIST 2017 library is made available by the authors.

On the contrary, RASSP~\cite{rassp} is a~significantly more sofisticated spectra prediction 
model. In the first step, it computes embedding of atom features with a~graph neural network.
The features are then combined with representation of all possible subformulae
using an attention head.
The outputs of the attention are further passed to a~feed-forward network together with
the subformulae embedding to compute probability distribution over the possible subformulae.
This is used in the final step of weighted sum of subformulae spectra.
RASSP reports superior accuracy compared to NIST.
However, due to the complicated embedding of subformulae and atoms, the current implementation
is rather restricted to eight most common elements only, and rather limited number of possible subformulae.

Transformer~\cite{attention} is a~recently emerging architecture of deep neural network 
used in many successful applications in natural language processing.
Specifically BART~\cite{bart} is an implementation of almost vanilla transformer
in the HuggingFace~\cite{hf} framework.

\section{Materials and Methods}

\begin{figure}
\begin{center}
\includegraphics[width=\hsize]{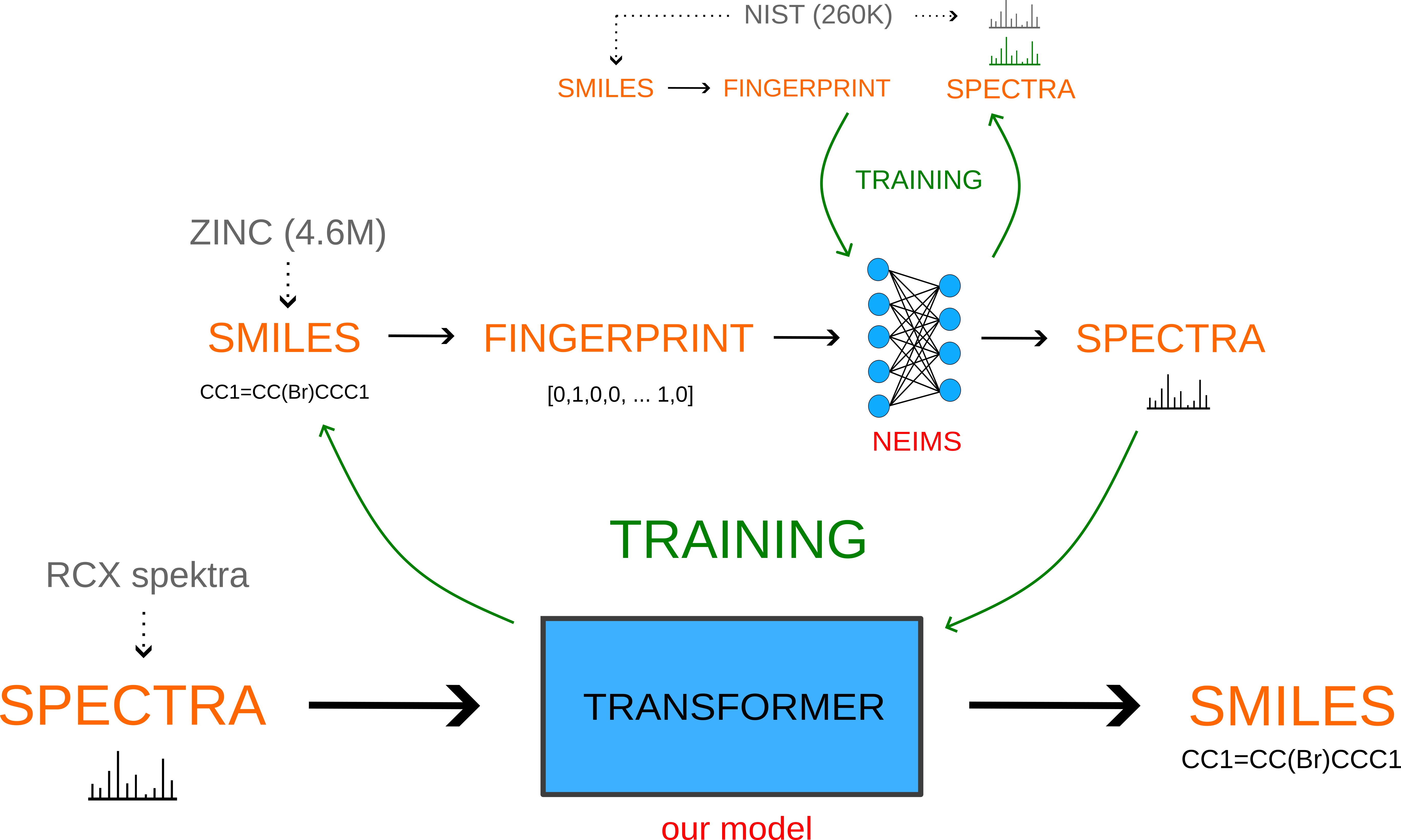}
\end{center}
\caption{Overview of the method}
\label{f:schema}
\end{figure}

\subsection{Overview}

Figure~\ref{f:schema} shows the main components and data flows of our method.
We start with a~set of known mass spectra with annotation (at least SMILES of the compound).
These data are used to train a~\emph{forward model} which predicts a~EI mass spectrum
given a~structural formula (SMILES) on input.
Despite the more data are used to train such model, the better, 
it turns out that reasonably accurate models can be built using data that are actually available.

In the next step, much larger ($100$--$1000\times$) number of structural formulae 
are fed to the forward model to make it predict their spectra.
These predicted spectra are used in turn to train a~much larger \emph{reverse model},
which predicts structural formula given a~mass spectrun on its input.

Finally, the reverse model is fine-tuned with the set of real spectra, originally used to
train the forward model.

These steps are described in detail in the following sections.

\subsection{Train and test data preparation}
\label{train-test}

According to our knowledge, the most comprehensive database of EI-MS spectra is the NIST library 
We used our purchased copy of the~2020 release.
The library contains spectra of 306,643 unique compounds and additional 43,774 replicate spectra (independent
measurements of the same compounds).
The data are provided in a~proprietary format, we used a~semi-automated procedure%
\footnote{\url{https://drchrispook.com/2021/04/08/convert-nist-mainlib-and-replib-ei-libraries-to-msp-format-to-annotate-gc-ms-data-with-ms-dial/}}
to convert it to an open MSP format.
Unfortunately, the conversion is not error-prone%
\footnote{For the time being, we did not dig into detail, approx.~5~\% loss is acceptable for our purposes.}
, we end up with 290,312 records of unique compounds.

This set was reshuffled randomly and it was split further in 90:10 ratio into training and testing set.
We double checked there is no overlap to avoid compromising the test results by information leak.

\subsection{Train the forward model}
\label{train-forward}

The task of the forward model is spectra prediction given a~structural compound formula.
In our workflow, we train it with an available spectral library, and we use it in turn 
to generate large number of training spectra for the reverse model.
The candidate models are also briefly evaluated for accuracy.


The authors of NEIMS~\cite{neims} provide an optimally trained model for download, and it yields 
an expected average accuracy 
on our test data.
Therefore we use it in the further calculations as is.

RASSP~\cite{rassp} reports significantly superior accuracy of spectra prediction compared to previous methods.
However, its available implementation is still rather restricted and experimental. 
Besides the limit on size of the molecule -- 48 atoms, which is acceptable for our purposes,
it is restricted to eight elements (H, B, C, N, O, F, P, S) only, and the provide configuration
restricts the number of generated subformulae to 4096. 
A~trained model is provided by the authors as well, however, we were not able to achieve its reported accuracy.
Therefore, we attempted to re-train the model.
Given the constraints on ``acceptable'' molecule, our training set reduces to 112,709 spectra only.
We were able to achieve somewhat worse accuracy
than reported in~\cite{rassp}.

The RASSP restrictions on molecule size, set of elements etc.\ affect requirements on GPU memory significantly, the original results are reported
to be achieved on commodity GPUs with 12~GB memory only.
By trial and error, we end up with extending the element set with Si, Cl, Br, I;
all other elements occur in NIST library less than $500\times$, and we could afford omitting 
these records (Sect.~\ref{train-test}).
Adding the new elements required minor extensions of RASSP code in its final step
(mapping subformulae probabilities to actual spectra).
Further, we set the limit on subformulae number to 100,000, causing omission of only few
hundreds of NIST records.
On the other hand, we keep GPU memory usage safely below 40~GB (on single formula+spectrum processing), 
which fits into current high-end GPUs (NVidia A40 or A100), and multiple GPUs can be
used simultaneously to process bigger batches.
However, besides rather frequent crashes while training, the model trained in this way performs significantly worse.

Altogether,
we stick with the simpler NEIMS for the time being, the more promising RASSP approach will
require further work.
Also, due to incomparably simpler architecture, NEIMS is also considerably faster, which makes a~difference
on huge datasets that we need to process.

\subsection{Training data of the reverse model}

We aim at developing as universal model as possible, capable of identifying \emph{any} compound
that is detectable with GC-EI-MS, including, e.g., yet-unknown metabolites.
We start with a~compound selection ``2D-clean-annotated-druglike'' from ZINC15~\cite{zinc} database.
In our opinion, this selection matches compounds that are likely to really exist while it is not
biased, as e.g.~``in stock'' would be.
Such a~search outputs approx.~1.8 billion of compounds, out of which we pick a~random selection of 30 million;
this number, together with several dozens of peaks in a~typical generated spectrum,
was chosen to match the number of trainable parameters (360 million) of the reverse model according to the best practices.
We extract SMILES from the record, and make them canonical with standard RDKit%
\footnote{\url{https://www.rdkit.org/}} procedure.
Then, stereoisomers (meaningless for EI-MS) are removed and too long SMILES (over 100 characters) are dropped. 
Finally, the set is filtered not to contain any compound present in our NIST-based testing set (Sect.~\ref{train-test})
in order to prevent jeopardizing our end-to-end evaluation by information leak at this point.

Our forward model (NEIMS, currently) is applied on this approx.~4.6M set to produce predicted EI-MS spectra --
the principal training set or the reverse model.
The output spectra consisting of more than 200 peaks are dropped (too many peaks indicates poor NEIMS prediction typically),
as well as spectra with peaks of $m/z>500$ which are irrelevant for gas chromatography.
Technically, we also keep aside a~small testing set at this stage 
to be used in the auxiliary evaluation of the reverse model (Sect.~\ref{eval-bart-neims}).

\subsection{Reverse model architecture}
\label{reverse-arch}

We aim at the original transformer~\cite{attention} model architecture.
From the available robust implementation we chose BART~\cite{bart} from the HuggingFace framework~\cite{hf}.
Despite of its different purpose, the architecture is almost the same as the original transformer suitable
for our purpose.
We keep the ``big'' BART configuration of 12 transformer layers of 16 attention heads each for both encoder and decoder,
and we use 4,096 as the width of feed-forward networks connecting transformer layers; other 
hyperparameters (including activation and loss functions) are preserved.
Altogether, the size of the model is approx.~360 millions of trainable parameters.
Those are initialized randomly, due to entirely different domains of available pre-trained
models we do not expect any benefit from using them.

Embedding of the input spectra is twofold. The $m/z$ values, rounded to the nearest integer eventually, become the input tokens,
yielding fairly small vocabulary size (500).
Intensity values are transformed logarithmically and they are binned into only 10 distinct values.
Thus, we still distinguish between large and small peaks, keep some resolution even among the small ones, while 
lowering the importance of intensity compared to $m/z$.
Spectra encoded in this way are padded to the fixed length of 200 peaks, and accompanying masks are generated
to exclude the padding from being considered by the encoder input layers.
The intensities are fed into the model through the original ``position'' channel, and they are added to the
main input in early stage according to the transformer architecture.

For the embedding of structural formulae (SMILES) the \emph{bbpe} tokenizer
adapted from HuggingFace GPT-2 implementation is used.
We took a~sample of 1~million SMILES from our ZINC15 set, and we trained the tokenizer on them independently of the
main model. This preprocessing yields a~table of 1,233 distinct tokens, generated according to their occurrence in 
the training set only, without any chemical meaning (and in most cases, they do not represent feasible substructures);
this is still fine because any valid SMILES can be represented by a~sequence of such tokens, and the approach
proved to be functional with natural languages.
On the contrary to the spectral input, the embedding of SMILES (to a~1,024-dimensional vector space)
is a~fully trainable part of the model.

Training of the model is done in a~standard way of encoder-decoder transformer. The input batch contains of pairs of
tokenized and padded spectra and SMILES. The SMILES part is further rearranged into a~sequence of prefixes (growing by one token)
and target ``labels'' -- the expected output tokens.  
Categorical cross-entropy loss is computed for all labels and the error is back-propagated through the whole model in a~usual way.
For the time being, we do not use any learning rate control or early stopping strategies; instead we evaluate model behaviour
manually.

Inference (prediction) of the transformer involves feeding the spectrum into the encoder part to compute its hidden state, and
auto-regressive repeated calls to the decoder. 
The latter can follow different strategies, and it yields non-deterministic results due to random sampling 
of the output token distributions.
We chose the recent \emph{contrastive search}~\cite{contrastive} which is reported to outperform previous approaches.

\subsection{Improvements with reinforcement learning}
Reinforcement learning (RL) has been a promising technique for fine-tuning transformer-based models in different research areas, such as natural language generation or audio captioning \cite{openai2023gpt4, peng-etal-2020-reducing, xu2022sjtu}. Using RL, researchers can optimize the parameters of transformers through a trial-and-error process, where the model receives feedback signals, such as rewards or penalties, based on its performance in a specific task. This approach is particularly useful when the objective function is complex, non-differentiable, or unavailable. 
In the case of the mass spectra to SMILES translation task, we are limited by the token-based language modelling loss function. Even though this type of loss is highly convenient for generating a string similar to the original SMILES representation of a given molecule, the similarity of actual molecules being represented can differ.
We suggest fine-tuning the model on the NIST dataset of measured spectra utilizing reinforcement learning. The idea is to use a molecular similarity as a reward function and optimize the model directly by this metric.

\begin{figure}
\begin{center}
\includegraphics[width=.67\hsize]{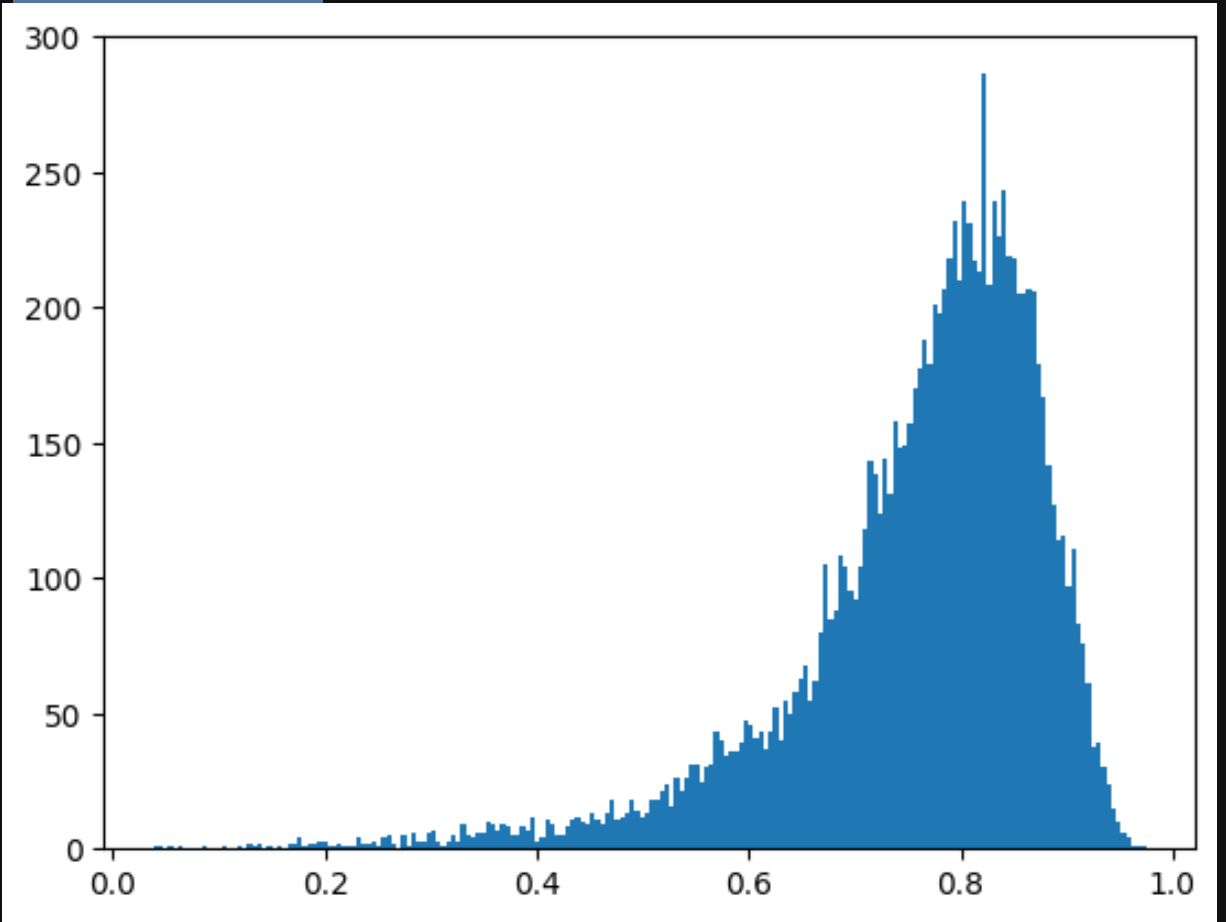}
\caption{Histogram of spectra similarities ($\textrm{DP}_{1,0.5}$) of NEIMS prediction on test set}
\label{f:neims-hist}
\end{center}
\end{figure}

\begin{figure}
\begin{center}
\includegraphics[width=.67\hsize]{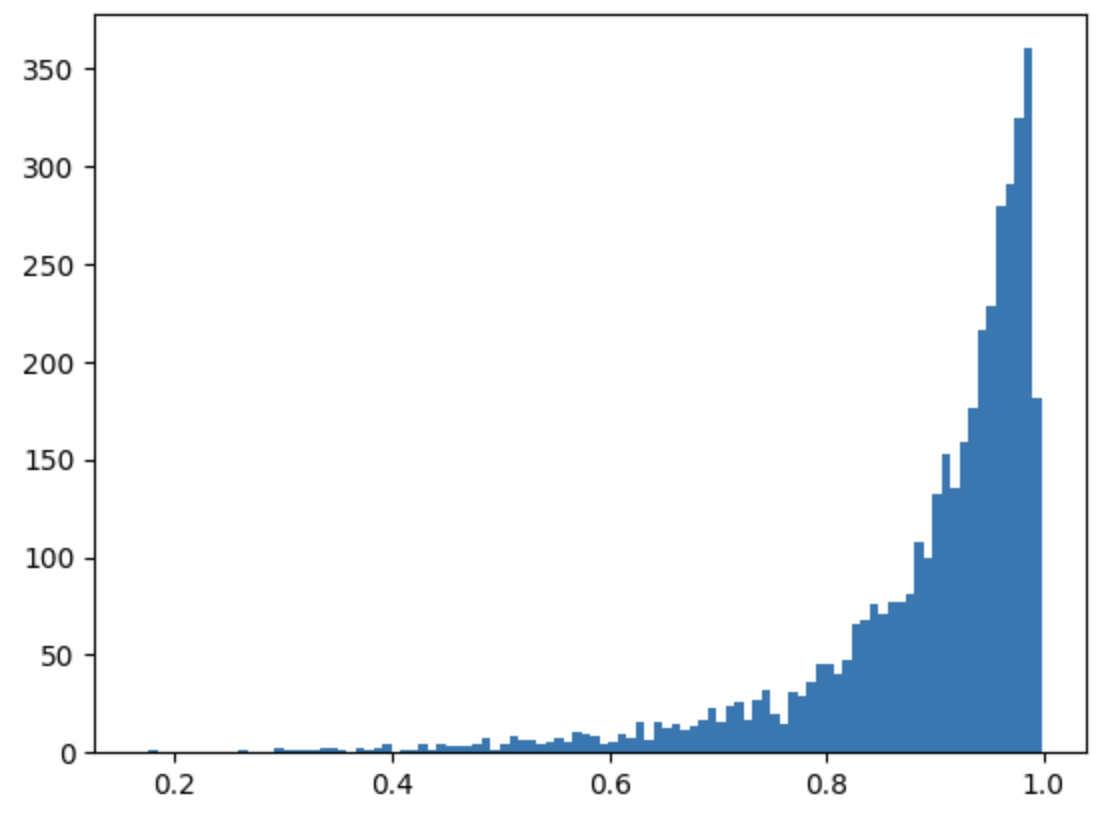}
\caption{Histogram of spectra similarities ($\textrm{DP}_{1,0.5}$) of RASSP prediction on test set}
\label{f:rassp-hist}
\end{center}
\end{figure}

\section{Results}

\subsection{Forward model accuracy}

One of the driving hypotheses of this work is that the accuracy of the forward model needn't be superior,
while the reverse model is still able to extract useful information from the forward model outputs.
A~similar phenomenon is leveraged successfully by large language models, which also work with imperfect but huge
test corpuses on input, still being able to derive general language rules out of them.
Therefore, we measure the accuracy of the forward models as an auxiliary indicator only,
to assess that the forward model ``still performs reasonably'',
according to expectations derived from relevant literature.

The model is evaluated using the metric to compare spectra $S_p, S_r$:
$$
\textrm{DP}_{a,b}(S_p, S_r)  = \frac {\sum_k m_k^a I_{pk}^b \times m_k^aI_{rk}^b}
		 {\big\|\sum_k(m_k^a I_{pk}^b)^2\big\| \big\|\sum_k(m_k^a I_{rk}^b)^2\big\|}
$$
which maps to regular dot product with $(a,b) = (1,0.5)$ and e.g.\ Stein dot product with $(a,b)=(3,0.6)$.
Figure~\ref{f:neims-hist} shows histogram of spectra similarities of NEIMS prediction over our test set,
the average $\textrm{DP}_{1,0.5}$ is 0.76.

RASSP is a~more sofisticated and more promising model, we achieved much better $\textrm{DP}_{1,0.5}=0.89$
and a~similar histogram (Fig.~\ref{f:rassp-hist}) leans to right apparently.
However, the current RASSP implementation is too restricted (see Sect.~\ref{train-forward}) to be used 
with our data.

\subsection{Evaluation of reverse model}

Because of the generative decoder of out model (Sect.~\ref{reverse-arch})
which involves rather random sampling of token probability distributions,
its outputs are not deterministic anymore, and it makes the evaluation of the reverse model
performance less straightforward.

For each spectrum in the test set (see subsections below), we let the reverse
model generate 10 candidate outputs (SMILES).  Because the actual
output of the transformer is a~softmax probability distribution on tokens (SMILES fragments),
we are able to calculate an overall probability that this output is generated by the model;
the outputs are sorted according to this probability, then.

For each of those, we compute their fingerprints, as well as for the reference
SMILES, and compare the fingerprints with Tanimoto index (common approach to
compare SMILES, e.g.~\cite{ohagan}), and we sort the results according to this similarity.
We draw histograms of the similarity of this best result over the testing set
(with the background that this statistics mimics the intuition of the human operator,
who would choose from the 10 candidates),
as well as we compute average similarities of the best results,
from the 3 top results, and from all unique results in the original 10 most probable outputs.

\begin{figure}
\begin{center}
\includegraphics[width=.67\hsize]{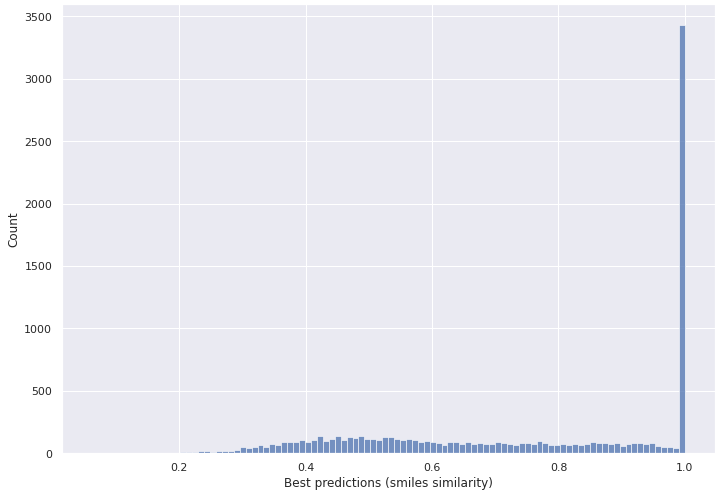}
\end{center}
\caption{Histogram of SMILES (fingerprint) similarities of the best result on 
spectra generated with forward model}
\label{f:result1}
\end{figure}

\subsubsection{Outputs of forward model}
\label{eval-bart-neims}


The first evaluation is done on spectra generated by the forward model
on SMILES from the test set.
In this way we evaluate the performance of the reverse model only, 
not biased by eventual inaccuracy of the forward model. 

The similarity histogram is shown in Fig.~\ref{f:result1}.
In~34~\% cases there is a~perfect match, the distribution of the remaining cases is more or less
flat.
The average similarity of the best prediction is 0.74, which still represents fairly good structural
match, 0.70 for three best, and 0.58 for all unique results out of the 10 top-ranked candidates.

\begin{figure}
\begin{center}
\includegraphics[width=.67\hsize]{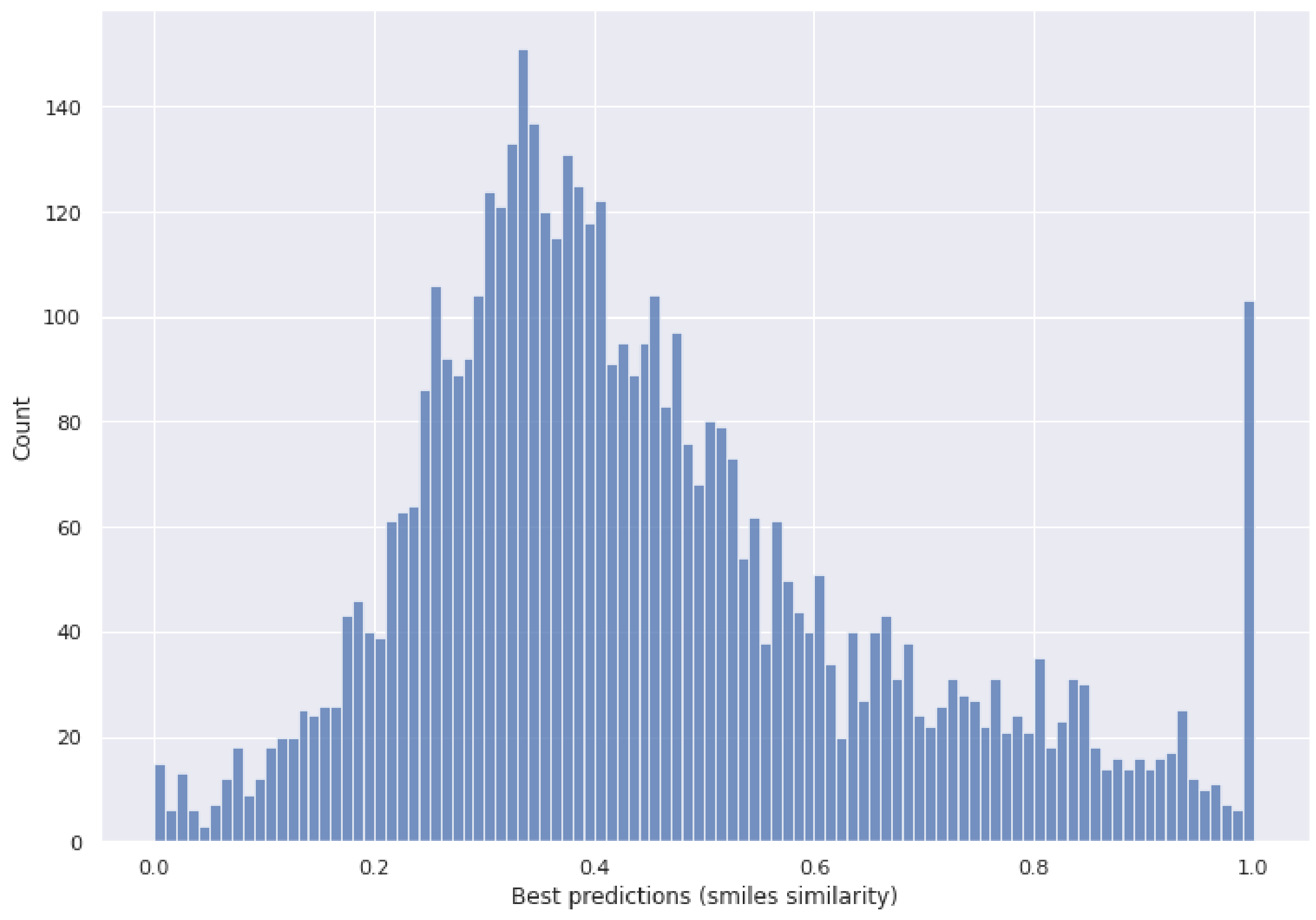}
\end{center}
\caption{Histogram of SMILES (fingerprint) similarities of the best result on 
experimental spectra (NIST)}
\label{f:result3}
\end{figure}

\subsubsection{Experimental spectra}

The model as a~whole is evaluated on the original experimental spectra from the NIST
library. Therefore it covers the overall performance of both forward and reverse models.
The similarity histogram is shown in Fig.~\ref{f:result1}
Compared to the previous experiment, the accuracy is significantly worse. 
The number of perfect matches drops to 5~\% only, most of the cases fall in
the similarity interval $[0.2, 0.6]$ which is not a~very good structural match anymore.
The average similarity of the best prediction drops to 0.45, 0.41 for the best three results,
and only 0.34 for all unique predictions.

\section{Discussion}

The general conclusion of our work is that the BART transformer architecture 
(enhanced with the positional encoding we use to feed in information on intensity)
is capable of performing the ``reverse'' mapping, i.e.\ prediction of molecular structure
from its GC-EI mass spectrum, using the spectrum only (no further information on precursor etc.).
The model generates ranked list of candidates, which can be assessed manually, filtered
further according to predicted vs.~measured chromatography retention index~\cite{deeprei}.

Unlike similar models~\cite{massgenie} we use a~pre-trained tokenizer on the SMILES side, 
resulting in a~much richer (1,233) token space w.r.t.~straightforward use of element symbols, bonds, 
parentheses and digits. 
Consequently, our model virtually never outputs invalid SMILES.

From comparison of the model with known models from NLP domain and further discussions
with experts we get convinced that increasing the model size (approx.~doubling the number
of trainable parameters) will help deriving more knowledge from the train data, and it will
improve the overall accuracy.


Apparently, the overall quality of transformer training data (outputs of the forward model)
have significant impact on the reverse model performance. 
This findings provide better lower bound to the original hypothesis,
the average spectra similarity of $\textrm{DP}=0.76$ we are able to achieve with 
NEIMS is not sufficient for this purposes.
This is the main reason the accuracy of our model is still not competitive
with the most similar work, MassGenie~\cite{massgenie}. 
However, exact comparison of the results is difficult,
they work on different data modality (MS/MS), and they use far more relaxed evaluation method
-- 10-fold larger set of candidates, and cherrypicking from among several metrics,
whichever gives the best score.

The recently published work on RASSP gives good hope, it reports DP over 0.95, and we are
almost able to reproduce this result.
However, the current implementation of RASSP is rather restricted on the size
and variability of the input molecule, and it must be extended for our purposes.

Finally, inspired by ``reinforcement learning from human feedback'' approach in the NLP tasks,
we expect to achieve further improvements in this way.

\section*{Implementation}
Complete source code of the reverse model implementation, together with data preparation and 
evaluation, is available at \url{https://github.com/hejjack/gc-ms_bart}.
Our extensions and evaluation of RASSP, as a~candidate forward model, are available at 
\url{https://github.com/ljocha/rassp-public/}.
The NEIMS forward model is used unchanged.
In order to reproduce our results, the NIST EI-MS library%
\footnote{\url{https://www.nist.gov/programs-projects/nist20-updates-nist-tandem-and-electron-ionization-spectral-libraries}}
must be provided for training.
We are not authorized to redistribute this data, neither the derived models.

\section*{Acknowledgement}

This work was supported by FR~CESNET, grant no.~679R1/2021.
Computational resources were provided by the e-INFRA CZ project (ID:90140),
supported by the Ministry of Education, Youth and Sports of the Czech Republic.

\bibliographystyle{amsplain}
\bibliography{bibliography}

\end{document}